# The Number of Terms and Documents for Pseudo-Relevant Feedback for Ad-hoc Information Retrieval

Mohammed El Amine Abderrahim[1], Saïd Benameur[2], Mohammed Alaeddine Abderrahim[3]

[1] University of Tlemcen,
Laboratory of Arabic Natural Language Processing
BP 230 Chetouane, Algeria

[2] University of Tlemcen,
Laboratory of Arabic Natural Language Processing
BP 230 Chetouane, Algeria

[3] University of Tlemcen,
Laboratory of Arabic Natural Language Processing
BP 230 Chetouane, Algeria

**Abstract**

In Information Retrieval System (IRS), the Automatic Relevance Feedback (ARF) is a query reformulation technique that modifies the initial one without the user intervention. It is applied mainly through the addition of terms coming from the external resources such as the ontologies and or the results of the current research.

In this context we are mainly interested in the local analysis technique for the ARF in ad-hoc IRS on Arabic documents. In this article, we have examined the impact of the variation of the two parameters implied in this technique, that is to say, the number of the documents «D» and the number of terms «T», on an Arabic IRS performance.

The experimentation, carried out on an Arabic corpus text, enables us to deduce that there are queries which are not easily improvable with the query reformulation. In addition, the success of the ARF is due mainly to the selection of a sufficient number of documents D and to the extraction of a very reduced set of relevant terms T for retrieval.

## 1. Introduction

In order to reduce the distance between the system's relevance and that of the user, an IRS can lead the user to a useful formulation of his needs. The suggested solutions can appear in various approaches i.e.: the Query Reformulation (QR), the reestablishment of documents, the combining in between all the results set of different IRS and or the integration of the user's profile in the retrieval information process.

In this article, we are basically interested in the QR. The approaches for the latter are numerous and can be classified according to the resources used in three great classes. (see Fig. 1):

- The use of the external resources: this consists of using the external resources as the ontologies or the thesaurus to find other terms similar to those in the initial query.
- The global analysis: this approach aims at analyzing the set of documents collection, so that we can extract the pertinent terms to be added to the initial query. So, we get two techniques that are: similarity thesaurus and the statistical thesaurus [5].
- The local analysis: the documents coming back to a query after demand are analyzed so that we can extract other pertinent terms that are able to extend the query. The studies applied in [3, 5, 11, 15, 16, 19, 21, 22, 27, 28, 29, 31] show that conversely to the global analysis, the local one is simpler to be realized and allows an improvement of IRS performances. Two techniques are proposed for the local analysis in the literature [5]:
  - The local clustering: this consists of constructing a matrix of association that quantifies the correlation relations of terms got from the set of documents that were returned in response to the initial query. According to the method of construction of the correlation relations, we notice three types of clusters: association clusters, metric clusters and scalar clusters. We are interested to this technique. So, we are going to develop this technique and we implement the first type of clusters i.e.: association clusters.
  - The local context analysis: it consists of using the concepts instead of the keywords to represent the document [16].

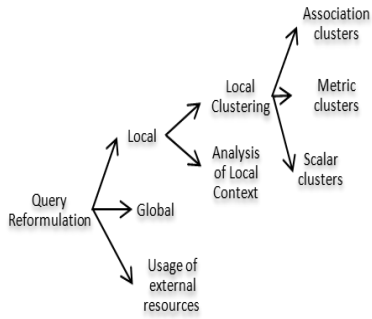

Fig. 1 The approaches for the query reformulation.

We distinguish two manners for the QR, the first one is based on an automatic process, it occurs without the intervention of the user. On the contrary, the second one is based on the interactive process between the IRS and the user. This is an interactive query reformulation (IQR). The experiments applied on behalf of this second manner have shown that it can allow the improvement of the results precision, all the same its efficiency is linked strongly to the attitudes of the users and their way of judging the documents pertinence [5, 9, 14, 20, 22, 26].

In order to avoid the heaviness of the judgment step of the documents relevance, this task has been automatized, and so the IRS considers the «D» first documents retrieved initially as pertinent. By these documents «D» the system establishes a list of «T» terms for the query expansion. This new form of pertinence reinjection is called blind, pseudo or ad-hoc (PRF). According to [9] and [12], this approach allowed to improve the results in comparison to the approach of global analysis. In fact, the attempts applied in the area of the campaign of evaluation TREC4 for the system SMART [12] and the work done by [11] assure that the technique of PRF leads to a gain in precision of responses of about 10%. In this article and in the area of the PRF, we suggest to examine the influence of the variation of «D» and «T» on the performances of an Arabic IRS. After this introduction, we are going to describe the PRF technique and we exhibit our experiment and we will discuss about the result.

## 2. The Relevance Feedback (RF)

The RF is a technique that consists of modifying the initial query of the user by adding some terms got from the list of documents retrieved in the IRS. It is based on three steps (see Fig. 2):

- The samples: it consists of selecting a set of «D» documents (samples) among the returned ones by the IRS and judged as pertinent.
- The extraction of evidences: it consists of establishing the list of «T» terms judged pertinent for the expansion of the query.
- The rewriting of the query: It consists of enriching the query with terms found in the previous step.

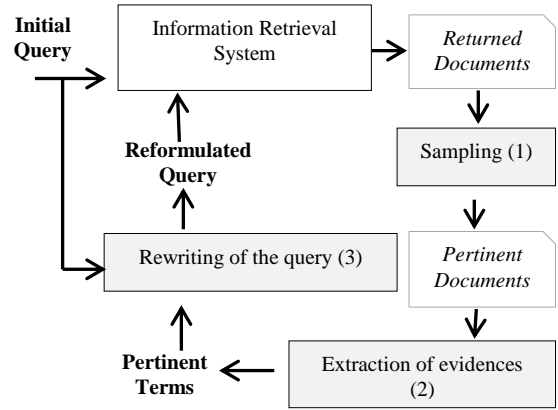

Fig. 2 The three phases of the relevance feedback.

The PRF is an approach for the RF that lies on the automatic sampling, otherwise, instead of judging explicitly the documents, we suppose that the «D» first documents are relevant.

The main problem with the PRF technique is summed up in the determination of two parameters «D» and «T». In the majority part of the work, these two parameters are chosen arbitrary, for instance:

- [23] has used: D=80 and T=10.
- [25] has used: D=20 and T=10.
- [21] has used: D= (5, 10, 25, 50, 75 and 100) and T= (10, 25, 50, 75 and 100).
- [16] has used: D= (5, 10, 20, 30, 50 and 100) and T= 70.
- [10] and [13] have used: D= (6, 10, 20, 25 et 30) and T= (1, 2, 3, … and 100).
- [12] proposed to use a «D» between 5 and 10 and recommend «D»=5 as an optimum value.
- [27] have used D=10 and T=30.
- [30] have used D=20.
- [4] and [2] have used: D=5.

On the other hand all the works that point at the effect analysis of the two parameters «D» and «T» on the IRS performance are not numerous. We find as examples:

- The experiment of [7, 8, 9] consists of analyzing the average precision of the IRS for all the combinations of the values: «D» and «T» between 1 and 100 (10 000 combinations). A list of 50 queries of the TREC8 was used with a collection of documents composed of newspapers articles. The results of the experiment have shown that there was an improvement in the average precision of the IRS for the values of «D» between 8 and 16 and the values of «T» between 7 and 42. It should be noted that for the same experiment but with a different collection of documents (TREC9 WT) composed of web pages, [9] has not obtained an improvement in the average precision. To explain that we find out that the web pages do not contain a pure content. They contain mainly some different objects that have nothing to deal with the content of the page such as: hyperlinks, decorations (images, animations, logos… etc.), interaction and other information (copyrights, information of contacts…). As for the matching of the queries, all these informations are considered as a noise and consequently degrading the performances of the IRS. This kind of content requires some approaches as the one used by [23] who has obtained some gains of performances at the order of 27%. The work set by [7, 8, 9] have led to two important results:
- The values D=15 and T=13 give the best value of average precision of the IRS.
- The estimated parameters (D, T) of one query are not obligatory suitable for another.

- In [10] many factors on which an IRS depends were studied. In the context of the PRF the variation of «D» and «T» was also examined. These experiments have not allowed to extract a conclusion concerning the estimated values for both «D» and «T». The final objective of the experiment was rather to analyze and try to understand the interaction between the IRS and the topics.
- [30] shows that the use of 10 terms leads to an improvement of more than half of the queries. However, there is no optimal and fixed number of terms for all queries.

The works concerning the RF evaluation for the Arabic texts are not enough, we find as examples:
- The researches done in [18] have shown that the manual RF by weighting the query terms have led to an improvement of the Arabic IRS performances (recall and precision). For his experiment, [18] has used a corpus of 242 documents and a list of 9 queries.
- The researches done by [6] linked to the query expansion with terms got from a thesaurus showed an improvement in the recall of the Arabic IRS. We notice that the corpus used was the Quran.
- The work achieved by [17] showed that the use of a thesaurus ameliorates considerably (18%) the performances of an Arabic IRS. It has also shown that the use of the roots is more effective than the use of schemes for the Arabic texts indexation.
- The work realized by [24] on the query expansion using an ontology in the field of law, and WordNet showed an improvement in the performances of the IRS.
- The work achieved by [1] on the query expansion using Arabic WordNet and a morphological analyzer showed an improvement in the recall but not the precision of the Arabic IRS.
- The work achieved by [2] on the strategy evaluation of the local PRF for the Arabic texts showed an improvement in the performances of the IRS.

In the context of this article, we are interested in the evaluation of the PRF technique for Arabic texts and particularly in the study of the two parameters (D, T) variation.

## 3. Experimentation

As for us, we are going to investigate the influence of the variation of the two parameters «D» and «T» involved in the PRF technique in Arabic IRS. For such a deal, we are going to vary «D» and «T» from 1 to 20 and for each combination (D,T) (with D=1, 20 and T=1, 20; 400 combinations) we are going to compare the different results obtained by the different systems (runs).

As for our experiment, we have used a corpus of Arabic texts in different fields. This corpus has been not used in an official evaluation campaign. The table 1 sums up the principal characteristics of this corpus.

Table 1: The main features of the corpus used

| Number of text files | 22 000 |
|---|---|
| Fields | Health, sports, politics, sciences, religion, astronomy, nutrition, law, tales, family |
| Size | 180 MB |
| Number of words | 17 000 000 |
| Number of different words | 612 650 |

All the operations concerning the indexation and the interrogation of the documents collection are realized using the API Lucene version 3.0 ( http ://lucene. apache. org). On the other hand, the process of PRF is developed in Java language. It implements the local clustering technique (see Fig. 3) to extract the most pertinent «T» terms that serve in the reformulation of the initial query. The table 2 shows examples of queries before and after reformulation.

Table 2: Examples of queries before and after RF.

| N° | Query before RF | Query after RF |
|---|---|---|
| 1 | أضرار التدخين (damage of smoking) | طب (Medicine)، مدخن (Smoker)، عام (year)، موقع (location)، مواضيع (topics)، هنا (here)، سلب (robbed)، اثار (traces)، عادة (habit)، علاج (treatment)، علم (science)، جديد (new)، مقال (article)، امراض (disease)، اخر (another)، زوار (visitor)، عين (eye)، جمع (collect)، دراسة (study)، قائمة ( list)، مصاب(infected) |
| 2 | سهم مالي (monetary action) | اسهم(Shares)، تداول(trading)، بنك (bank)، سوق(market)، مال (money)، عام(Land) |
| 3 | مدار الارض (Earth's orbit) | ارض( earth )، قمر (moon)، شمس (sun)، مدار ( orbit )، فلك (orbit)، رئيس(Chairman)، موقع(location) |

```
// Algorithm of the experimentation
Begin
 For each D (D=1, 20) do
  For each T (T=1, 20) do
   For each query qi (i=1, 50) do
    1- Interrogation of the collection of documents
    2- Sampling : select the «D» first returned
       documents: D_F
    3- Extraction of evidences
       - Construct the matrix of local association
         (term - term) from the set of distinct terms
         of D_F : S⃗   with each element
```

$$S_{u,v} = \sum_{d_j \in D_F} f_{S_{u,j}} \times f_{S_{v,j}}$$

$f_{S_{u,j}}$ : represents the term frequency

$S_u$ in the document $d_j$

$S_{u,v}$ : expresses the correlation

between $u$ and $v$

- For each term $t$ of $q_i$ extract its local association clustering $C_i$ set from the «T» highest values $S_{u,v}$ ($v \neq u$) of the $u^{th}$ line of $\vec{S}$

4- **Rewriting of the query** : construct the new query
  $q_{new} = q_i \cup C_i$
5- **End**

Fig. 3 Algorithm evaluation of the PRF (local clustering technique).

For each combination (Di, Tj) (i = 1, 20, j = 1, 20), the results of different queries are written in different files. For example: the number of documents found, the number of relevant documents found, the precision at 5, 10, 20, 100 and 1000 documents (P@5, P@10, P@20, P@100, P@1000) and the average precision.

## 4. Analysis and discussion of the results

To understand the effect of the variation of D and T on each query, we have established various measures that are mainly based on the comparison of results before and after enrichment.
For a given query, three cases can arise:
- Improvement (+): All precisions (at 11 points of recall) before are lower than those after. In other words, the curve (recall / precision) after is over before.
- No improvement (-): is the inverse of the previous case. The curve (recall / precision) before is over after.
- No decision (X): for some precisions, there is an improvement but for others there is no improvement. In other words, there is an intersection of the two curves (recall / precision).

### 4.1 Comparison based on the number of improved queries

We recorded in Table 3 for each value of D (respectively T) the average number of queries improved.

Table 3: (a) the average number of improved queries in accordance with D. (b) the average number of improved queries in accordance with T.

| D | Average Number of queries (+) | D | Average Number of queries (+) | T | Average Number of queries (+) | T | Average Number of queries (+) |
|---|---|---|---|---|---|---|---|
| 1 | 2,30 | 11 | 4,25 | 1 | 3,65 | 11 | 2,80 |
| 2 | 1,70 | 12 | 3,90 | 2 | 3,90 | 12 | 2,95 |
| 3 | 4,40 | 13 | 3,90 | 3 | 3,70 | 13 | 3,10 |
| 4 | 3,45 | 14 | 3,90 | 4 | 3,55 | 14 | 3,00 |
| 5 | 3,70 | 15 | 4,05 | 5 | 4,25 | 15 | 3,20 |
| 6 | 3,55 | 16 | 3,30 | 6 | 3,60 | 16 | 3,15 |
| 7 | 2,80 | 17 | 2,95 | 7 | 3,75 | 17 | 3,05 |
| 8 | 2,75 | 18 | 2,90 | 8 | 2,85 | 18 | 3,10 |
| 9 | 3,35 | 19 | 2,90 | 9 | 3,15 | 19 | 3,20 |
| 10 | 3,55 | 20 | 2,80 | 10 | 3,25 | 20 | 3,20 |

(a)                                                                  (b)

We note in Table 3 (a) that the average number of improved queries tends to increase when the value of D increases and vice versa in Table 3 (b) it decreases when the value of T increases.

## 4.2 Comparison based on the number of improved tests (runs)

We recorded in Table 4 the number of tests that led to an improvement (+) of the number of queries with the values of the corresponding D and T.

Table 4: The number of tests that led to an improvement (+) depending on the number of improved queries with the values of D and T.

| Number of queries (+) | Number of tests | D | T |
|---|---|---|---|
| 0 (0%) | 2 (0.5%) | 2 | 1,4 |
| 1 (2%) | 10 (2.5%) | 1 | 16,17,18,19,20 |
| 2 (4%) | 86 (21.5%) | 17,18,19, 20 | 8,9,10,11,12,13,14 |
| 3 (6%) | 108 (27%) | 5,6,7,8,9, 10 | 11,12,13,14,15,16 |
| 4 (8%) | 151 (37.75%) | 10,11,12,13, 14, 15 | 1,2,3,4,5,6, 7 |
| 5 (10%) | 40 (10%) | 3,4,11,13,15, 16 | 1,4,5,7 |
| 6 (12%) | 3 (0.75%) | 14,15,19 | 5,2 |

The analysis of the results of table 4 enables us to release the following remarks:
- In only two tests (D = 2, T = 1) and (D = 2, T = 4), no reformulated query has improves the IRS performance. This percentage is very low and we can confirm that the query reformulation improves the IRS performance.
- In three tests among 400 (0.75%), we achieved a maximal improvement of 12% (6 queries from 50).
- In 259 tests (151 108), about 65%, an improvement of 7% was recorded.
- In 191 tests (151 +40), about 48%, we achieved an improvement of 9%.
- Approximately 50% of the 151 systems (having shown an improvement of 8%) have a value of D between 10 and 15, and T between 1 and 7. This fact confirms the results obtained in the previous section (4.1), i.e., improving the performance of an Arabic IRS can be achieved by high values of D and low values of T.
- Approximately 50% of the 108 systems (having shown an improvement of 6%) have a value of D between 5 and 10, and T between 11 and 16.
- Approximately 50% of the 86 systems (having shown an improvement of 4%) have a value of D between 17 and 20, and T between 8 and 14.

In conclusion, we can say that the values {14, 15, 19} for D and {5, 2} for T can enable a maximum improvement of the IRS performance. Remember that these values are only valid for the corpus used in our experiments. We not pretend, in any case, a generalization of these results.

## 4.3 Comparison based on the number of improved queries

The table 5 shows for each query used in the experiment the number of systems with improvement indicator (+), no improvement (-) or no decision (X).

Table 5: The number of systems according to the three cases of query comparison (Improvement (+) No improvement (-), no decision (X)).

| N° Query | (+) | (-) | (X) | N° Query | (+) | (-) | (X) |
|---|---|---|---|---|---|---|---|
| 1 | 0 | 0 | 400 | 26 | 243 | 17 | 140 |
| 2 | 0 | 0 | 400 | 27 | 0 | 0 | 400 |
| 3 | 1 | 390 | 9 | 28 | 0 | 381 | 19 |
| 4 | 0 | 0 | 400 | 29 | 0 | 40 | 360 |
| 5 | 0 | 321 | 79 | 30 | 14 | 0 | 386 |
| 6 | 2 | 4 | 394 | 31 | 1 | 266 | 133 |
| 7 | 0 | 0 | 400 | 32 | 0 | 385 | 15 |
| 8 | 0 | 16 | 384 | 33 | 20 | 0 | 380 |
| 9 | 0 | 384 | 16 | 34 | 0 | 0 | 400 |
| 10 | 69 | 46 | 285 | 35 | 0 | 1 | 399 |
| 11 | 9 | 20 | 371 | 36 | 109 | 27 | 264 |
| 12 | 228 | 10 | 162 | 37 | 0 | 250 | 150 |
| 13 | 20 | 208 | 172 | 38 | 27 | 87 | 286 |
| 14 | 136 | 3 | 261 | 39 | 0 | 0 | 400 |
| 15 | 12 | 39 | 349 | 40 | 0 | 0 | 400 |
| 16 | 0 | 99 | 301 | 41 | 0 | 24 | 376 |
| 17 | 0 | 55 | 345 | 42 | 34 | 45 | 321 |
| 18 | 0 | 0 | 400 | 43 | 0 | 0 | 400 |
| 19 | 0 | 47 | 353 | 44 | 29 | 74 | 297 |
| 20 | 1 | 29 | 370 | 45 | 0 | 132 | 268 |
| 21 | 0 | 0 | 400 | 46 | 0 | 368 | 32 |
| 22 | 0 | 39 | 361 | 47 | 0 | 45 | 355 |
| 23 | 348 | 0 | 52 | 48 | 9 | 0 | 391 |
| 24 | 13 | 19 | 368 | 49 | 0 | 0 | 400 |
| 25 | 0 | 169 | 231 | 50 | 3 | 315 | 82 |

The analysis of the results of table 5 enables us to release the following remarks:
- Only 21 queries (42%) are concerned with the improvement (+).
- In 29 queries (58%), more than half, we did not obtain improvement. In other words, these queries are very difficult to improve with the query reformulation.

- In 12 queries (24%) we obtained indecision (X) in all tests (400 tests).
- From the point of view of the query domain, and knowing that we have worked with a text corpus of ten (10) different areas, we recorded at least one query improved by domain and two queries improved for seven domains (70%). Moreover, all queries have been improved for one domain only (law domain). Based on these results, we can conclude that there is no direct relationship between the query domain and the query reformulation.

4.4 Comparison on the basis of the analysis of the precision at 5 documents (P@5)

From the point of view of the P@5, the analysis of the results obtained has enabled us to deduce that:
- For 166 of the 400 systems (41.5%) the P@5 after is greater than the P@5 before, so we can confirm that the query reformulation improves the IRS performance. In addition, 202 of the 400 systems (50.5%) the P@5after is less than the P@5 before, so there is no performance improvement for the IRS. Table 6 shows the combinations of D and T for the best eight (8) P@5 obtained.

Table 6: T and D for the best P@5 obtained

| D | T | P@5 |
|---|---|-----|
| 3 | 3 | 0,660 |
| 3 | 4 | 0,656 |
| 3 | 2 | 0,644 |
| 18 | 2 | 0,644 |
| 5 | 2 | 0,644 |
| 17 | 2 | 0,640 |
| 18 | 8 | 0,640 |
| 18 | 13 | 0,640 |

The examination of the various values of D and T that allows for an improvement in P@5 are generally between 3 and 20 for D and between 2 and 20 for T. Also, we found that the increase in D (or T) does not necessarily imply an improvement in P@5. Moreover, if we fix the value of D, the increase in T does not consequently increase (improve) the P@5. The opposite is also true, that is to say, if we fix the value of T, the increase of D does not necessarily imply an increase (improvement) in P@5.

## 5. Conclusion

In IRS, automatic relevance feedback is a technique for reformulating the user query. It is based on a process composed of three steps: the sampling, the terms extraction and the rewriting of the initial query.

Two parameters are to be taken into account during the two first steps which are: the size of the sample D and the number of terms T to be extracted for the query expansion.
In this paper, an experiment was conducted on a corpus of Arabic text in order to study the effect of the variation of the two parameters in question (D and T) on the overall performance of the Arabic IRS.
The aim is not to propose optimum values for D and T and even less, to generalize the results on Arabic SRI.
The experiment allows us to deduce that there are queries that are hardly improvable with the query reformulation. Moreover, the success of the PRF is mainly due to the selection of a sufficient number of documents and the extraction of a reduced set of relevant terms for the search.
The results obtained allow us on the one hand, to study the correlation between D and T. On the other hand, opening the way to test the same technique with different types of corpora in order to provide the best possible settings for each query.

**Mohammed El Amine Abderrahim** is a research teacher at the University of Tlemcen, Algeria. His research interests are natural language processing, information retrieval, information extraction, databases and data mining. Med El Amine has a Magister in computer science from UST Oran, Algeria, and a Doctorate in computer science from the University of Tlemcen, Algeria. He is a member of the Laboratory of Arabic Natural Language Processing, university of Tlemcen.

**Saïd Benameur** is a research teacher at the University of Tlemcen, Algeria. His research interests are natural language processing, applied linguistics and translation. Saïd has a Magister in linguistics from the University of Tlemcen, Algeria. He is a Doctorate candidate and a member of the Laboratory of Arabic Natural Language Processing in the University of Tlemcen.

**Mohammed Alaeddine Abderrahim** is a research teacher at the University of Tiaret, Algeria. His research interests are natural language processing, information retrieval, information extraction, data mining and ontology. Med Alaeddine has a Magister in computer science from the University of Tlemcen, Algeria. He is a Doctorate candidate and a member of the Laboratory of Arabic Natural Language Processing in the University of Tlemcen.